\PassOptionsToPackage{utf8}{inputenc}
\documentclass[final]{bioinfo}
\copyrightyear{2019} \pubyear{2019}
\usepackage{xcolor}
\usepackage{hyperref}

\hypersetup{colorlinks=true, linkcolor=blue}

\access{Volume 36, Issue 1, 1 January 2020, Pages 212–220}

\begin{document}
\firstpage{1}

\subtitle{Population genetics}

\title[Mycorrhiza: Genotype Assignment using Phylogenetic Networks]{Mycorrhiza: Genotype Assignment using Phylogenetic Networks}
\author[Georges-Filteau \textit{et~al}.]{Jeremy Georges-Filteau\,$^{\text{\sfb 1,}*}$, Richard C. Hamelin\,$^{\text{\sfb 2}}$, Mathieu Blanchette\,$^{\text{\sfb 1,}*}$}
\address{$^{\text{\sf 1}}$School of Computer Science, McGill University, Montreal, Canada \\
$^{\text{\sf 2}}$Department of Forest and Conservation Sciences, The University of British Columbia, Vancouver, Canada}

\corresp{$^\ast$To whom correspondence should be addressed.}

\history{Received on 23 July 2018; revised on 03 May 2019; accepted on 06 June 2019, published on 14 June 2019}
\editor{Associate Editor: Russell Schwartz}

\abstract{\textbf{Motivation:} The genotype assignment problem consists of predicting, from the genotype of an individual, which of a known set of populations it originated from. The problem arises in a variety of contexts, including wildlife forensics, invasive species detection, and biodiversity monitoring. Existing approaches perform well under ideal conditions but are sensitive to a variety of common violations of the assumptions they rely on.\\
\textbf{Results:} In this paper, we introduce Mycorrhiza, a machine learning approach for the genotype assignment problem. Our algorithm makes use of phylogenetic networks to engineer features that encode the evolutionary relationships among samples.  Those features are then used as input to a Random Forests classifier. The classification accuracy was assessed on multiple published empirical SNP, microsatellite or consensus sequence datasets with wide ranges of size, geographical distribution and population structure and on simulated datasets. It compared favorably against widely used assessment tests or mixture analysis methods such as STRUCTURE and Admixture, and against another machine-learning based approach using PCA for dimensionality reduction. Mycorrhiza yields particularly significant gains on datasets with a large average FST or deviation from the Hardy Weinberg equilibrium. Moreover, the phylogenetic network approach estimates mixture proportions similarly well or more accurately than STRUCTURE and Admixture for most simulation cases. \\
\textbf{Availability:} Mycorrhiza is released as an easy to use open-source python package at \href{github.com/jgeofil/mycorrhiza}{\textcolor{blue}{github.com/jgeofil/mycorrhiza}}.\\
\textbf{Contact:} \href{jeremy.geo@gmail.com}{\textcolor{blue}{jeremy.georges-filteau@mail.mcgill.ca}}  and \href{blanchem@cs.mcgill.ca}{\textcolor{blue}{blanchem@cs.mcgill.ca}}\\
\textbf{Supplementary information:} Supplementary data are available at \textit{Bioinformatics} online.
\href{http://bit.ly/mycobio}{\textcolor{blue}{https://doi.org/10.1093/bioinformatics/btz476}}
}

\maketitle

\section{Introduction}

Assignment methods are a group of closely related methods that use genetic information to determine the population membership of individuals from a given species. For this purpose, the term “population” generally refers to a group of individuals in close geographical proximity whose probability of interbreeding is higher than that of interbreeding with other groups. In one version of the problem, called the assignment test, one aims to estimate the probabilities that a multilocus genotype of unknown origin came from each of a fixed set of known populations. This is equivalent to the classification problem in machine learning. In another version, called genetic mixture analysis or genetic stock identification, the objective is to estimate both mixture proportions and posterior source probabilities for each individual. In this paper we present and evaluate a new machine learning algorithm for genetic assignment based in part on phylogenetic networks.\\
Assignment methods have been used for a variety of applications, including wildlife forensics (Larraín et al., 2014; Schwartz and Karl, 2008; Glover et al., 2009; Lorenzini et al., 2011; Millions and Swanson, 2006), understanding migratory patterns and geographical boundaries for conservation efforts (Stewart et al., 2013) and the identification of hybrid individuals for the management of invasive species (Ibañez-Justicia et al., 2017; Johansson et al., 2018; Larraín et al., 2018; Michalecka et al., 2018; Dauphinais et al., 2018)). Despite their wide range of applications in a variety of fields and a number of well know software tools implementing various algorithms (see below), little consensus exists about their use in classical supervised classification problems. \\
Assignment methods have mostly been implemented with frequentist, maximum likelihood or Bayesian analysis algorithms (Manel et al., 2005). Most  approaches assume that the loci used as features are at Hardy-Weinberg equilibrium and in linkage equilibrium (Manel et al., 2005; Kalinowski, 2011; Pritchard et al., 2000-6; Falush et al., 2003). In real-world populations, these assumptions are rarely fully satisfied for the total set of available loci. Efforts have been made to overcome these limitations by relying on genetic distances rather than allelic frequencies, without considerable success (Cornuet et al., 1999).\\
The widely used STRUCTURE program (Pritchard et al., 2000-6; Novembre, 2016) is a model-based, Bayesian clustering method for explicitly inferring population structure and probabilistically assigning individuals to  populations (Pritchard et al., 2000-6; Falush et al., 2003). Although originally introduced as an unsupervised clustering approach, the method has since been enhanced with a semi-supervised model in which some of the individuals can be pre-assigned to their known population of origin (Jonathan K. Pritchard, Xiaoquan Wen, Daniel Falush, 2010; Porras-Hurtado et al., 2013). This capability can also be used to emulate what is know as supervised learning the machine learning field. Unfortunately, STRUCTURE suffers from its high computational complexity when applied to large SNP datasets, in which case running time can be on the order of days or even weeks (Raj et al., 2014). As an alternative, FastSTRUCTURE improves the computational efficiency of STRUCTURE using a variational Bayesian framework (Raj et al., 2014). However, unlike STRUCTURE, it cannot account for linkage disequilibrium.\\
Admixture (Alexander et al., 2009) is another tool for maximum likelihood estimation of individual ancestry. It is based on the same statistical model as STRUCTURE, but optimized with a block relaxation algorithm (Zhou et al., 2011). According to the authors, it is as accurate as STRUCTURE, but with the added advantage of being much more computationally efficient. In practice however, STRUCTURE tends to be more accurate than Admixture as it can partially account for linkage disequilibrium between markers (Falush et al., 2003).\\
Bayesian clustering methods in general come with many known limitations. They are, for example, known to lose their ability to detect subpopulations at very low levels of population differentiation (Latch et al., 2006). Studies have also shown that when the parameter K is smaller than the actual number of populations, STRUCTURE can produce clusters inconsistent with their evolutionary history (Kalinowski, 2011). This can also happen when the size of populations is unbalanced (Neophytou, 2014; Kalinowski, 2011; Puechmaille, 2016; Wang, 2017).\\
Methods based on Principal component analysis (PCA) or other multivariate analysis methods have long been used as fast and efficient tools to analyse structure in genomics data sets. When used simply as a visualization tool, these methods do not however, allow for straightforward interpretation of individual ancestry from the low-dimensional projection they produce. Alternative clustering methods, such as discriminant analysis of principal components (DAPC) (Jombart et al., 2010), have been developed with the aim of providing fast and flexible exploratory tools that produce easily interpretable results. However, these methods either only allow for hard clustering or provide questionable admixture results with soft-clustering (Lee et al., 2009).\\
A number of R packages have been developed with the goal of offering machine learning solutions for genomics. The package Adegenet was developed with the aim of bridging the gap between multivariate data analysis solutions and genomics packages by implementing a number of clustering algorithms such as snapclust (Beugin et al., 2018) and discriminant analysis of principal components (Jombart, 2008; Jombart and Ahmed, 2011; Beugin et al., 2018). The package can calculate a number of population statistics and perform spatial genetics analyses. However, snapclust, the algorithm for population assignment implemented in the package, is also based on the same assumptions about Hardy-Weinberg equilibrium as Bayesian methods.\\
Overall, no existing software addresses all of the shortcomings and limitations mentioned above. In this paper, we set out to develop a new method for genotype assignment and mixture analysis rooted in machine learning principles that would address these problems, while keeping in mind and taking advantage of the phylogenetic structure present in genomics datasets.
\subsection{Phylogenetic networks as feature engineering}
Phylogenetic networks have been introduced to capture and represent non tree-like evolution (Morrison, 2014a; Huson and Bryant, 2006; Bandelt et al., 1995; Hendy and Penny, 1993; Bandelt and Dress, 1992). By allowing for hybrid nodes, or reticulations, these are better suited to account for more complex evolutionary events such as hybridization, horizontal gene transfer and recombination. In explicit phylogenetic networks, each internal node represents a hypothetical ancestor. On the contrary, this is not necessarily the case in implicit networks.  One such type of implicit network, the split network, can be interpreted as a combinatorial generalization of unrooted phylogenetic trees (Huson and Bryant, 2006).\\
Both phylogenetic trees and networks are composed of a set of splits, referred to as a "split system". A split $S$, of the form $S = A|B$, is a bipartition of the set of taxa (or specimens) into two nonempty subsets $A$ and $B$ (Huson et al., 2010). The distinction between a split system describing a phylogenetic tree and one describing a phylogenetic network is in the set of compatibility rules that the set of splits must satisfy (Semple et al., 2010). A split network, or phylogenetic network, is a graphical representation of an underlying split system. An example phylogenetic network, along with the associated split system represented by a binary matrix is shown in Fig. 1. Splits for which one of the subsets is of size one (e.g. split 2, in blue) are said to be trivial. A trivial splits represent genetic divergence that is unique to its corresponding sample.\\
In the graphical representation of a split system, each split is represented by one or more parallel edges. In most cases a weight, corresponding to a dissimilarity measure is associated to each split. The evolutionary distance between two taxa is thus the sum of the weights of all splits that place these taxa in different subsets.

\begin{figure}[]
    \includegraphics[width=1.0\linewidth]{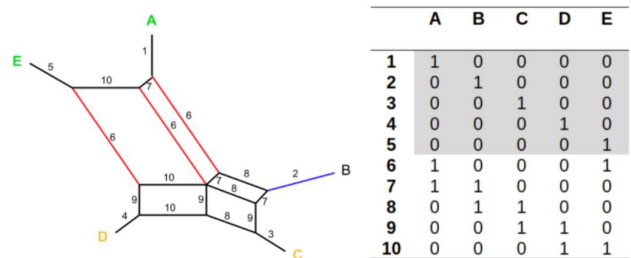}
    \caption{Example phylogenetic network and corresponding split system. The taxa are denoted by the letters A to E and splits are numbered from 1 to 10. Each unique split is represented by one line (trivial split, e.g. split 2, in blue) or multiple parallel lines (e.g. split 6, in red). In the split system (right), each taxon is placed on either side of the splits, as indicated by a binary flag. Trivial splits are shaded in grey. In this example, taxa are colored according to their population of origin (A and E in green, D and C in yellow), while B is from an unknown population we could want to predict.}\label{fig:01}
\end{figure}
Split networks can be inferred from pairwise distances between the set of taxa. The NeighborNet algorithm will produce a split system that perfectly fits the input distance matrix, provided it is circularly decomposable (Bryant et al., 2007; Levy and Pachter, 2011). The running time of NeighborNet is O(n3), making it is suitable for large data sets.\\
Implicit phylogenetic networks have been used  for visualisation, data mining and exploratory data analysis as a means of fuzzy clustering (Morrison, 2014b; Yu et al., 2014) and explicit phylogenetic networks are used to infer evolutionary histories in the presence of reticulation events (Yu et al., 2014; Yu and Nakhleh, 2015; Solís-Lemus and Ané, 2016). Interestingly, there also exists a few recent cases of network inference being used as a data transformation step in algorithms (Fioravanti et al., 2018; Chernomor et al., 2016; Wicke and Fischer, 2017; Volkmann et al., 2014). This makes sense considering that the splits of a circular split system are linearly independent and very little information is lost when resolving a distance matrix (Bryant and Dress, 2007). We therefore propose the use of split system decomposition as a feature space reduction method. For our purpose, the reduced feature set is then inputted to a Random Forest classifier for geographical origin prediction. We named this approach Mycorrhiza, a term defined as the symbiotic association between a fungal network and the roots of host plants.
\begin{methods}
\section{Methods}
\subsection{Mycorrhiza}
Mycorrhiza is composed of two main steps. In the first step, a phylogenetic split system is inferred from the genotype data of all individuals (of both unknown and known origin). In the second step, the placement in the split system of individuals of known origin, along with their categorical label (corresponding to their geographical origin), is used to train a Random Forest classifier. Following this, the trained model is used to make predictions from the split placement of individuals of unknown origin.\\
\textbf{Step 1: Split system inference}\\
To produce a split system, pairwise genetic distances are calculated for all $N$ individuals over $M$ single site loci. For SNP and sequence data, the Jukes-Cantor distance was calculated with MEGA-CC (Kumar et al., 2012). For microsatellite data, the distance was calculated as the number of loci with different copy number. Following this, a phylogenetic split system is built from the pairwise distances. Based on this matrix, the NeighborNet program from the SplitsTree4 (version 4.14.6) package (Huson and Bryant, 2006) is used to obtain a circular phylogenetic split system. Only splits with a weight above $10^{-6}$ are considered. Although both the labeled and unlabeled examples are used to build the split system, this is done in an unsupervised fashion, without knowledge of the categorical population labels. Finally, binary feature vectors of length $S - N$ are extracted from the S dimensional split system (discarding the $N$ trivial splits), to be used in the learning and prediction step. Each individual is represented by a binary vector corresponding to its placement on either side of each of the non-trivial splits. In Fig. 1, sample A, for example, would be represented as with feature vector $[1,1,0,0,0]$. Trivial splits are ignored because they do not have any discriminatory power. In practice, a split system can contain up to a few thousand splits, depending on the number of samples and the complexity of the population structure.\\
\textbf{Step 2: Training and predictions}\\
The feature vectors built based on the split system in step 1 are used as input to a Random Forest classifier. The scikit-learn implementation of a Random Forest was used for this purpose with default parameter settings, except for the number of estimators which was set to 60 (see Discussion). Alternate families of classifiers such as fully connected neural networks were also evaluated, but they proved less accurate in this context.\\
The trained model is then used to make predictions for unlabeled examples, based on their split vector. The output of Mycorrhiza corresponds to the probability that a sample belongs to each of the  populations, as estimated by the random forest model. If the desired output is a hard classification to a single population, each sample is attributed the population label that has the highest probability. Alternately, the probability distribution can be interpreted as the individual's mixture proportions over the K populations.
\subsection{Partitioned Mycorrhiza}
In machine learning, ensemble methods, which combine the output of multiple classifiers trained on different subsets of the features, have been shown to reduce generalization error. This is, in fact, the strategy behind Random Forests. We thus developed a variant of Mycorrhiza called Partitioned Mycorrhiza in which $P$ different Mycorrhiza predictions models are trained, each on feature vectors produced from disjoint subsets of loci. In other words, $P$ split networks are inferred from disjoint subsets of loci. The final output is obtained by averaging the  predictions. By default, $P$ is set to $10$.
\subsection{PCA variant}
For the PCA variant, the first step of the algorithm is replaced by a standard PCA analysis on the distance matrix. Here, the placement of a sample in the D-dimensional principal component space yields its feature vector. The feature vector is then used as input to a Random Forest classifier in the same way it was for Mycorrhiza. We used the scikit-learn implementation of PCA for this purpose. After evaluating different choices of values for the number of components $D$, it became clear that no improvement in accuracy was obtained beyond $D=50$, and that the results were quite robust to that parameter. We thus set $D = min(N, 50)$ for all data sets. If needed, this hyper-parameter could easily be optimized on a per-dataset basis, e.g. using standard cross-validation procedures.
\subsection{Number of loci and partitioning parameters}
To investigate the effect of feature set size on classification accuracy, the number of loci used as input was varied by randomly downsampling to the desired number of loci m. Assignment accuracy was evaluated with 5-fold cross-validation. This was repeated $5$ times and the results were averaged. For SNP data, $m$ was varied from $50$ to $M$ by powers of $2$. For microsatellite data, $m$ was varied from $2$ to $M$ by increments of $2$. For sequence data, the values of $m$ were chosen more arbitrarily, depending on $M$ which varied greatly between datasets.\\
For Partitioned Mycorrhiza, the same procedure was applied to evaluate the effect of the number of partitions on classification accuracy. The $m$ loci were further divided into $p$ subsets, that were used to build $p$ split placement feature sets, and classification accuracy was averaged over five runs for every combination of $m$ and $p$. For SNP data, $p$ was set to $1$, $2$, $10$, $50$, $100$ or $500$.  For microsatellite data, $p$ was set to every value between $1$ and $M$.
\subsection{Simulated data and mixture proportions}

Simulation data sets containing 5 distinct populations were produced with SLiM 3.2.1 (Haller, 2019) according to three common migration models (Jombart et al., 2010; Rodríguez-Ramilo et al., 2009). Each population contained $500$ diploid individuals with $100000$ single site loci. Simulations were run for $3000$ generations, with migration starting at generation $2000$. The mutation rate was set $10^{-7}$ and the recombination rate was set to $10^{-8}$. In the island model, the rate of migration is the same for all pairs of populations. Simulations for the island model were done setting the migration rate to $0.0001$, $0.0003$, $0.0004$ or $0.0005$. In the hierarchical model, populations are divided into groups of $2$ and $3$ populations, with the migration rate between pairs of populations in the same group being higher than the migration rate between pairs in different groups. Simulations for the hierarchical model were done setting the inner migration rate to $0.001$ and the outer migration rate to $0.0001$, $0.000125$ or $0.00015$. In the stepping stone model, populations are organised in a linear fashion, with migration happening between immediate neighbors only. Simulations for the stepping stone model were done setting the migration rate to $0.0002$, $0.0005$, $0.001$, $0.002$, $0.003$ or $0.004$. At the end of each simulation, $50$ individuals were randomly sampled per population to produce the final data sets.

\subsection{Implementation and software package}
Mycorrhiza is released as an easy to use open-source python package on GitHub at \underline{github.com/jgeofil/mycorrhiza}. The package depends on the SplitsTree software (Huson, 1998). Data can be inputted in a number of commonly used formats in population genetics such as STRUCTURE, Genepop and many others. Partitioning parameters can be used in their default setting or optimized with predefined procedures. Common cross-validation methods are implemented to calculate classification accuracy, but static training and testing sets can also be used. Estimated mixture proportions can be outputted as a text file or as a figure. 
\end{methods}
\section{Results}
We analysed several empirical datasets of different types (SNP, microsatellite, sequence) and ploidy. SNP datasets were filtered for a minor allele frequency of $>=0.05$ and disallowing any sites for which data is missing for certain individuals. Summary statistics for these datasets are presented in Table 1 (see Suppl. File 2 for reference). Mycorrhiza, Partitioned Mycorrhiza, PCA+RF, Partitioned PCA+RF, STRUCTURE and Admixture were applied in turn to each of these datasets and to each of the simulated datasets to evaluate classification accuracy (See Suppl. Methods).\\
The mean squared mixture error was also calculated for each method on the simulated datasets as $\frac{1}{N}\sum_{i}^{N}\sum_{j}^{K}(x_{ij}^{R} - x_{ij}^{E})^2$, where $N$ is the number of individuals, $K$ is the number of populations, $X^{R}$ are the real mixture proportions and $X^{E}$ are the estimated mixture proportions. Simulated datasets contained on average $556$ SNPs and deviation from the Hardy-Weinberg equilibrium was lower than $0.007$ in all cases. 

\begin{table*}[!tbp]
\processtable{Summary statistics of the datasets on which assignment methods were tested.\label{Tab:01}}
{\centering\begin{tabular}{@{}llllllllll@{}}
\toprule
 & & & & & & \multicolumn{2}{c}{\#Sam/pop} & &\\ 
\cline{7-8}
Dataset & Type & Ploidy & \#Loci & \#Pop & \#Sam & min & max & FST & FSTd\\
\midrule
\textit{Arabidopsis thaliana} (1001genomes.org) & SNP & D & 458 075 & 10 & 979 & 28 & 243 & 0.15 & 0.24\\
Brown Rat (Puckett et al., 2016) & SNP & D & 32 127 & 8 & 185 & 12 & 40 & 0.11 & 0.32\\
Gypsy Moth (Picq et al., 2018) & SNP & D & 2 327 & 8 & 90 & 10 & 12 & 0.37 & 0.56\\
Human (The 1000 Genomes Project Consortium, 2015) & SNP & D & 530 973 & 26 & 780 & 30 & 30 & 0.10 & 0.14\\
Rice (McCouch et al., 2016) & SNP & D & 458 475 & 20 & 740 & 20 & 50 & 0.16 & 0.29\\
\textit{Septoria musiva }(Sakalidis et al., 2016) & SNP & H & 519 848 & 8 & 83 & 7 & 19 & NA & 0.31\\
Asian Ladybird (Lombaert et al., 2014) & STR & D & 18 & 6 & 1318 & 87 & 501 & 0.03 & 0.05\\
\textit{Mycosphaerella fijiensis} (Robert et al., 2012) & STR & H & 21 & 21 & 678 & 12 & 66 & NA & 0.52\\
Oriental Fruit Moth (Kirk et al., 2013) & STR & D & 13 & 16 & 376 & 8 & 72 & 0.19 & 0.29\\
Yellow Fever Mosquito (Brown et al., 2011) & STR & D & 12 & 13 & 1152 & 30 & 185 & 0.16 & 0.23\\
Barnacle (Wrange et al., 2016) & SEQ & S & 96 & 12 & 434 & 26 & 57 & NA & 0.14\\
Ebola (Pickett et al., 2012) & SEQ & P & 16 817 & 3 & 794 & 239 & 300 & NA & 0.09\\
HIV (Foley, 2017) & SEQ & S & 3 191 & 5 & 628 & 28 & 150 & NA & 0.19\\
Seabird tick (Dietrich et al., 2014) & SEQ & S & 141 & 7 & 432 & 5 & 131 & NA & 0.62\\
\botrule
\end{tabular}}{D=diploid, H=haploid, P=polyploid, S=consensus sequence of diploid organism}
\end{table*}

\begin{figure}[!tbp]
    \includegraphics[width=\linewidth]{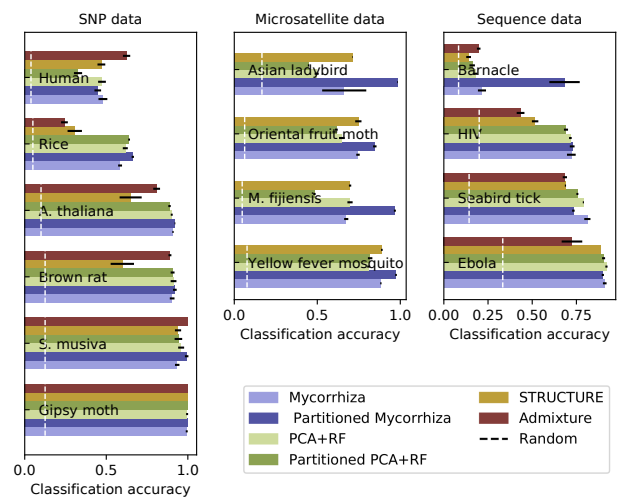}
    \caption{Classification accuracy estimated using 5-fold cross-validation for the tested assignment methods. Standard error of the mean calculated over 5 repetitions.}\label{fig:02}
\end{figure}

\subsection{Assignment accuracy}
Figure 2 shows the accuracies obtained by each tool on each of the empirical datasets. Overall, Mycorrhiza and Partitioned Mycorrhiza, both of which were run with the same set of default parameters for each dataset, correctly assigned individuals to their populations of origin with greater accuracy than both STRUCTURE and Admixture on the majority of tested empirical datasets. Furthermore, when compared only to their PCA counterparts, Mycorrhiza and Partitioned Mycorrhiza perform similarly or better on all tested empirical datasets.\\
With SNP data, Partitioned Mycorrhiza provided a slight advantage over Mycorrhiza and both PCA counterparts on the \textit{A. thaliana}, Brown rat, \textit{S. musiva} and Rice datasets. Generally, STRUCTURE and Admixture attained lower accuracy, with the exception of the human dataset for which Admixture outperformed our methods. With microsatellite loci, Partitioned Mycorrhiza considerably outperformed all other methods on all tested empirical datasets, by a margin ranging from $9$ to $27\%$ in accuracy. Finally, with sequence data, Mycorrhiza, Partitioned Mycorrhiza, PCA+RF and Partitioned PCA+RF performed nearly equally well on most empirical datasets. The only exception is the Barnacle data, for which Partitioned Mycorrhiza provided a $39\%$ increase in accuracy over the second best method.\\
Figure 6 shows the accuracies obtained by each tool on each of the simulated datasets. Mycorhiza, PCA+RF and Admixture performed equally well on all simulated datasets, while STRUCTURE attained significantly lower accuracy than the three other methods.
\subsubsection{Number of loci  and partitioning parameters}
We evaluated the impact of the number of loci used as input for all methods by executing them on randomly downsampled empirical datasets (Fig. 3). With SNP data, the accuracy of Mycorrhiza and PCA+RF (either with or without partitioning), as well as Admixture gradually increases with the number of loci, generally plateauing after $10,000$ loci. Interestingly, the number of loci at which the plateau in accuracy is reached with SNP data seems to depend on number of populations present in the dataset.
Results are similar for empirical microsatellites and sequence data, although most datasets are too small to reach a plateau in accuracy, suggesting that the inclusion of additional loci would further improve classification performance. Note that we expect that non-random loci selection would allow reaching a similar accuracy with much fewer loci. Surprisingly, the results obtained with STRUCTURE were somewhat more erratic.\\
We also assessed the extent to which loci partitioning improves performance for Mycorrhiza and PCA+RF (Suppl. Fig. 1). Setting the number of partitions to between $10$ and $50$ provided a consistent but moderate improvement in accuracy on nearly all empirical SNP datasets, with the notable exception of the Human dataset, where best results were obtained without partitioning. Accuracy gains obtained by partitioning are most striking for the empirical microsatellite datasets, where a number of partition around 10, corresponding to partitions of only $1$ or $2$ microsatellites each, yields a $30$ to $50\%$ improvement in accuracy. Results for empirical sequence data were not as consistent and did not indicate strong trends, although setting the number of partitions to $10$ is near-optimal for all datasets except for the oriental fruit moth. Based on these results, the default number of partitions for Partitioned Mycorrhiza was set to $10$ for all data types.
\begin{figure}[!tbp]
    \includegraphics[width=\linewidth]{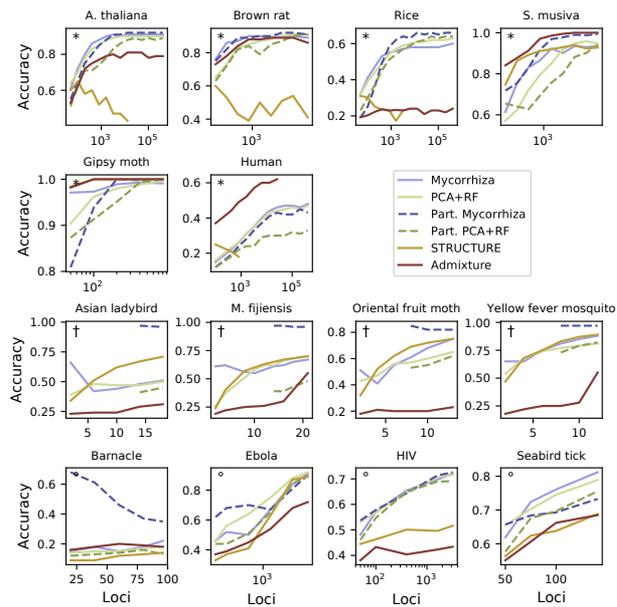}
    \caption{Accuracy versus the number of randomly selected loci for Mycorrhiza, Partitioned Mycorrhiza, PCA+RF, Partitioned PCA+RF, STRUCTURE and Admixture. Empirical SNP (*), microsatellite (†) and sequence ($\circ$) datasets. STRUCTURE did not terminate within the allocated time on the A thaliana, rice and human datasets when the number of loci, and model complexity, were too high. Partitioned Mycorrhiza and Partitioned PCA+RF can only be executed with 1 or more locus per partition on the microsatellite datasets. }\label{fig:03}
\end{figure}
\subsubsection{Impact of population structure statistics on prediction accuracy}
We next studied how population structure parameters (expected and observed heterozygosity, deviation from Hardy-Weinberg equilibrium and the fixation index of subpopulations; see Supplementary Methods) differentially impact prediction accuracy of Mycorrhiza compared to STRUCTURE and Admixture on empirical datasets (Fig. 4). As expected, population differentiation and levels of heterozygosity influence the ability of STRUCTURE and Admixture to capture population structure. Both variants of Mycorrhiza performed better than these two tools on empirical SNP datasets for which sub-populations had a high fixation index. The same trend can be observed when comparing the accuracy of these methods versus the level of population differentiation on simulated datasets (Fig. 6). The advantage of our methods  also increases proportionally to the level of heterozygosity of the total population and deviation from the Hardy-Weinberg equilibrium, but decreases proportionally to the average sub-population heterozygosity. With empirical microsatellite and sequence data, the trends are not as clear, possibly due to the smaller number of data sets being analyzed. SplitsTree provides a measure of fit describing to what extent the phylogenetic network produced accounts for observed distances between taxa. For Mycorrhiza, accuracy is generally higher when the fit is good, but Partionned Mycorrhiza seems to be less dependent on this measure (Figure 5). Finally, classification accuracy was generally higher for empirical datasets with a smaller number of populations, as it to be expected. For empirical SNP and sequence data, classification accuracy also generally increased with the number of loci. However no appreciable trend could be observed as a function of the number of samples.
\begin{figure}[!tbp]
    \includegraphics[width=\linewidth]{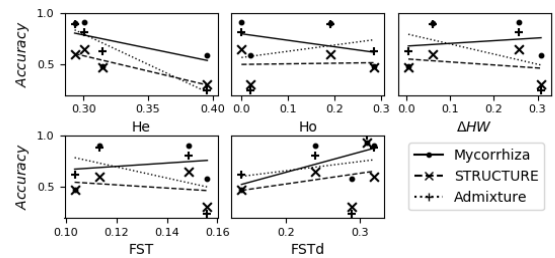}
    \caption{Assignment accuracy on empirical SNP datasets, as a function of the expected heterozygosity (He), observed heterozygosity (Ho) and deviation from the Hardy-Weinberg equilibrium ($\Delta$HW), and average population fixation index, calculated from heterozygosity (FST) and from genetic distances (FSTd).}\label{fig:04}
\end{figure}
\begin{figure}[!tbp]
    \includegraphics[width=\linewidth]{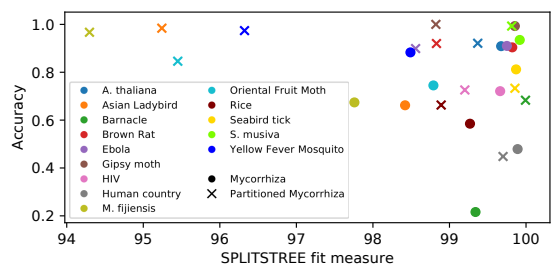}
    \caption{Classification accuracy on empirical data sets versus the SplisTree fit measure calculated over all loci.}\label{fig:05}
\end{figure}
\begin{figure*}[!tbp]
    \includegraphics[width=\linewidth]{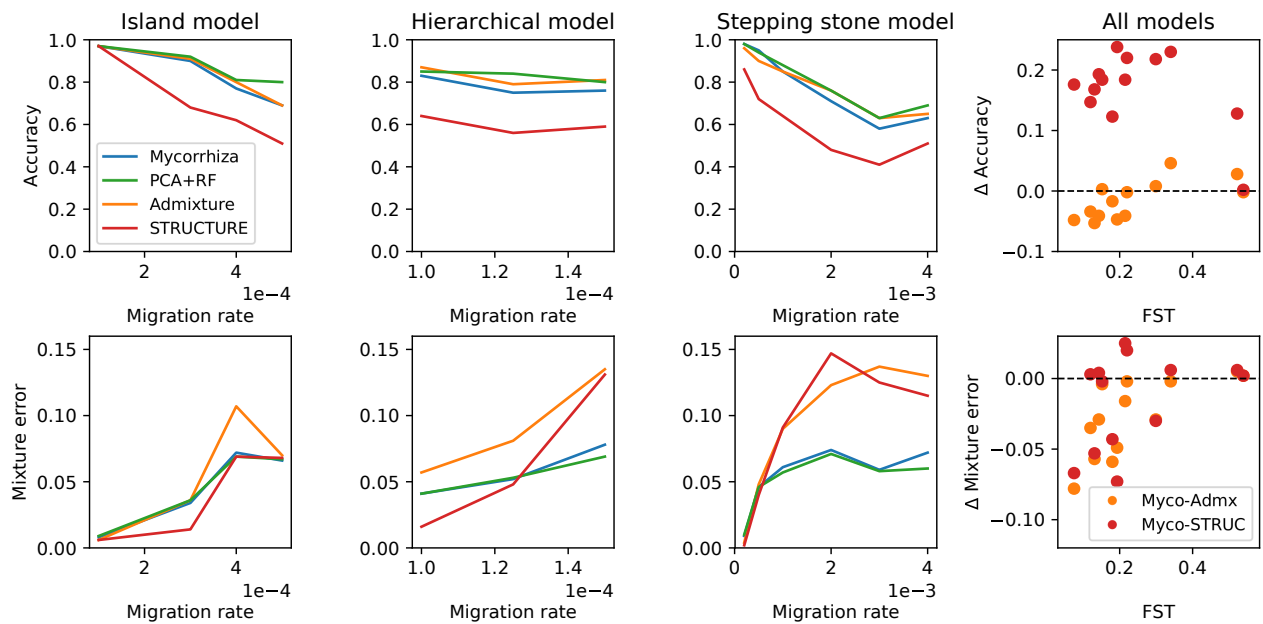}
    \caption{Simulated data sets. (Top left 3) Accuracy versus the migration rate. (Bottom left 3) Mean squared mixture error versus the migration rate. (Top right) Difference in accuracy between Mycorrhiza and Admixture/Structure versus the FST. (Bottom right) Mean squared mixture error between Mycorrhiza and Admixture/Structure versus the FST.}\label{fig:06}
\end{figure*}
\subsection{Estimation of mixture proportions}
Although Mycorrhiza and Partitioned Mycorrhiza was not specifically designed to estimate population mixture proportions, we observe that the population membership probabilities it outputs closely resemble those obtained with STRUCTURE, which is widely acknowledged to be accurate at estimating population mixture proportions. An example output representing mixture proportions obtained by all methods from the \textit{S. musiva} dataset with $800$ loci is shown in Suppl. Fig. 2. Notice also how PCA+RF appears to overestimate the contribution of secondary populations, even on datasets for which it performs well in terms of classification accuracy.\\
Figure 6 shows the mean squared mixture error obtained by each tool on each of the simulated datasets. For the island model datasets, all tested methods estimated mixture proportions similarly well. However, on the stepping stone and hierarchical datasets, Mycorrhiza and PCA+RF estimate mixture proportions more accurately when the rates of migration are higher.

\subsection{Running time}
The running time of Mycorrhiza with 5-fold cross-validation on each dataset is presented in Suppl. Fig. 3. Time was calculated with no partitions on a single core of a standard laptop computer (Intel(R) Core(TM) i5-7300HQ CPU @ 2.50GHz). All running times were under an hour and most were under $10$ minutes. They increase cubically with the number of samples, and linearly with the number of loci, although under approximately $10,000$ loci only the former has a significant impact. The number of populations in the datasets had no discernable impact on execution time. Partitioned Mycorrhiza with $P$ partitions takes approximately $P$ times longer to run than Mycorrhiza, but this could easily be parallelized. For comparison, the running time of STRUCTURE was at least five times larger, resulting in certain datasets taking several days, weeks or months to analyze. Admixture, is considerably faster than STRUCTURE and was comparable to Mycorrhiza in execution time (under an hour), although slightly slower.
\section{Discussion and conclusion}
We introduce Mycorrhiza, a machine learning method making use of phylogenetic networks to assign multilocus genotypes to their geographical origin. Mycorrhiza proved to be highly flexible, outperforming or equaling the other methods on all data types. Mycorrhiza is highly accurate, flexible and robust, outperforming or equaling in accuracy the most popular existing methods (STRUCTURE, Admixture) on a variety of empirical and simulated datasets based on SNPs, microsatellites, and sequence. The consistency of results produced by Mycorrhiza provides a considerable advantage for the method: default parameters will provide near-optimal results in almost all cases. Mycorrhiza’s accuracy improves gradually as the number of markers available increases, to eventually plateau, but never decrease. Furthermore, methods such as STRUCTURE and Admixture depend on assumptions about population structure that are often unmet in practice. Mycorrhiza is not directly dependent on these assumptions, and indeed, the data sets where populations exhibit high fixation index or strong deviation from Hardy-Weinberg equilibrium are those where the benefit of Mycorrhiza is most striking. Unsurprisingly, Admixture and, to a lesser extent, STRUCTURE performed poorly on empirical sequence data due to the fact that allelic frequencies of diploid or polyploid genotypes are impossible to estimate from a reduced sequence. Mycorrhiza, being based on distances rather than allelic frequencies, is seemingly less affected by this loss of information. Mycorrhiza is thus a tool that is simple and straightforward to use, requiring little or no parameter optimization, and exhibiting a high degree of robustness to the type of data at hand and the parameters of the population structure. Mycorrhiza would, in particular, be a tool of choice when dealing with geopolitically, rather than genetically, defined populations. For example, this is necessary when assignment test must be used in international trade discussions or to coordinate risk mitigation of invasive species between countries.\\
With the increasing throughput and affordability of DNA sequencing, genotype assignment problems are quickly becoming very large, both in terms of the number of samples and number of loci they contain. Computationally efficient algorithms are thus more necessary than ever. The inferior performance observed with STRUCTURE may be due in some cases to insufficient burn-in, although in nearly all cases the value reported by the program was below $0.2$, which according to the documentation suggests convergence. Increasing the burn-in period is thus unlikely to yield meaningful improvement and would be difficult to achieve due to already extremely long running time. Mycorrhiza’s running time, which is dominated by that of the $O(n^{3})$ NeighborNet algorithm (Bryant and Moulton, 2004), remains moderate (less than one hour) on even the largest datasets available today. Training our random forest model does not represent a significant computational burden. Moreover, improvements to the NeighborNet algorithm are published on a regular basis, including a recent report of 2-fold speed-up and 6-fold reduction in the memory footprint (Porter, 2018). Notably, unlike for Bayesian methods, the numbers of loci and of populations in the dataset have a negligible effect on running time. Overall, Mycorrhiza not only provides better classification accuracy in most datasets tested, but also reduces computation time considerably.\\
Although we limited our performance comparison to the two most widely used approaches in the field, some of the empirical datasets analyzed here had been previously analyzed by their authors using other tools. This provides additional points of comparison, albeit in a less rigorous framework. Picq et al. (2018) analyzed their gypsy moth data sets using DAPC (Jombart et al., 2010), reporting $100\%$ classification accuracy with all $2327$ SNPs, as well as with a selected set of $48$ SNPs. We produced similar results, obtaining perfect accuracy with as few as $200$ randomly selected SNPs ($99\%$ with $100$ randomly selected SNPs and $97\%$ with $50$). Results obtained for the yellow fever mosquito microsatellite data set are also consistent with the authors’ analyses. Using GeneClass2 (Piry et al., 2004) on $10$ non-problematic loci, the authors reported $87.7\%$ classification accuracy. In our hands, STRUCTURE obtained a similar performance ($88.9\%$), but Partitioned Mycorrhiza did much better, obtaining an accuracy of $98.4\%$. The low classification accuracy obtained for the Human data is somewhat surprising, but could be explained by the fact that this data set has the largest number of populations and a relatively low number of samples.\\
While already fast and accurate, several directions would be worth investigating to improve Mycorrhiza and further broaden its range of applications. Firstly, only implicit phylogenetic networks built using the NeighborNet algorithm were used in this study. Although this algorithm is known to produce well resolved and simple networks in most conditions, it would be interesting to study whether other network building algorithms could perform better, and under what conditions. However, extending the method to explicit phylogenetic networks, in which the notion of split is different, would require substantial modifications. Similarly, classifiers other than random forests, such as deep neural networks, may be advantageous for data sets with an extremely large number of samples; replacing our random forest predictor by such an alternative would be straightforward. Finally, Mycorrhiza currently makes no use of the weights assigned to splits by NeighborNet. Those weight could instead be used, e.g. to bias the sampling of the corresponding features by the random forest algorithm.\\ 
It is also worth noting that Mycorrhiza is applicable not only to genotypic data, but also to any other type of population-specific traits, including language or the morphological characteristics of artefacts. All that would be needed is the definition of a suitable phenotypic distance measure. In fact, phylogenetic networks have even been used to model relationships on a range of different data types, and Mycorrhiza may be applicable to these as well.\\
In conclusion, by combining sophisticated phylogenetic network reconstruction algorithms with machine learning approaches, Mycorrhiza represents a novel solution to the genotype assignment problem. Its accuracy, scalability, and most importantly its robustness to data types and sizes, and properties of underlying population structure, should make it an attractive solution for a wide array of population genetics researchers.
\section*{Acknowledgements}
This study was funded by Genome Canada, Genome British Columbia, Genome Quebec in support of the Large-Scale Applied Research Project in Natural Resources and the Environment BioSurveillance of Forest Alien Enemies (bioSAFE) project. The authors thank Monique Sakalidis for providing the SNP dataset for \textit{S. musiva}.

\clearemptydoublepage
\end{document}